# A classification-based adaptive segmentation pipeline: feasibility study using polycystic liver disease and metastases from colorectal cancer CT images

Peilong Wang[1], Timothy L. Kline[1], Andy D. Missert[1], Cole J. Cook[1], Matthew R. Callstrom[1], Alex Chan[1], Robert P. Hartman[1], Zachary S. Kelm[1], Panagiotis Korfiatis[1,*]

[1] Department of Radiology, Mayo Clinic, Rochester, MN, USA

[*] Corresponding author

Email: first author – Wang.Peilong@mayo.edu, corresponding author - Korfiatis.Panagiotis@mayo.edu

## Abstract

Automated segmentation tools often encounter accuracy and adaptability issues when applied to images of different pathology. The purpose of this study is to explore the feasibility of building a workflow to efficiently route images to specifically trained segmentation models. By implementing a deep learning classifier to automatically classify the images and route them to appropriate segmentation models, we hope that our workflow can segment the images with different pathology accurately. The data we used in this study are 350 CT images from patients affected by polycystic liver disease and 350 CT images from patients presenting with liver metastases from colorectal cancer. All images had the liver manually segmented by trained imaging analysts. Our proposed adaptive segmentation workflow achieved a statistically significant improvement for the task of total liver segmentation compared to the generic single segmentation model (non-parametric Wilcoxon signed rank test, n=100, p-value $\ll$ 0.001). This approach is applicable in a wide range of scenarios and should prove useful in clinical implementations of segmentation pipelines.

# Introduction

Segmenting organs and structures affected by pathologies is a complex and challenging task. The diverse shapes, sizes, and structures of organs and lesions introduce unique effects on radiologic appearance, which in turn affects the performance of segmentation models. While a trained model may demonstrate high accuracy in segmenting a specific anatomical region with a particular pathology, it often fails to produce satisfactory results when confronted with new underlying pathologies that differ from what the model was trained on. This issue necessitates the development of robust and adaptable segmentation approaches.

Researchers have explored strategies to address the segmentation of various pathologies. One approach is to improve the generic segmentation models [1-3], which aim to capture common features of organ structures or regions of interest across various pathologies. However, they often struggle to account for the distinct characteristics and variations exhibited by different pathologies, leading to suboptimal segmentation outcomes. Another avenue of exploration is the multi-task learning [4, 5], which involves training a model conducting segmentation and classification in a model simultaneously. This approach aims to enhance segmentation performance by leveraging shared knowledge and representations among different pathologies. However, multi-task learning can also be challenging, as manually adjusted multi-task weight is often imprecise and the imbalance issue of samples is difficult to address in the model.

In this study, we tackle the segmentation task of various pathology images in a two-step process. Initially, images are classified into distinct pathologies. Subsequently, images from each corresponding pathology category are segmented using a tailored segmentation model. These sequential steps constitute the innovative segmentation pipeline where images are automatically routed to the appropriate single-pathology-specific model for segmentation. In real-life clinical workflow, especially when a large number of cases are not available for various pathologies, this step-by-step approach can provide a practical and reliable solution. The data we used for this study are the polycystic liver disease (PLD) and liver metastases from colorectal cancer (MCC). PLD is a hereditary disorder characterized by the development of multiple fluid-filled cysts in the liver [6-9]. Colorectal cancer originates in the colon or rectum, with the liver being a common site for metastatic colorectal cancer (MCC) [10, 11]. Both PLD and MCC are medical conditions that can profoundly impact liver shape and function, leading to increased morbidity and mortality rates. Our pipeline is expected to achieve higher accuracies in the PLD and MCC liver segmentation. A generic segmentation model to segment both PLD and MCC pathologies simultaneously is also trained for the comparison with the pipeline.

## Materials and Methods

### Data

In our study, we constructed two datasets. A dataset comprising 350 CT scans of patients with PLD and another comprising 350 CT scans of patients with liver metastases from colorectal cancer (MCC). Each dataset was divided into 300 cases for training and validation, while 50 cases were reserved for testing. These 50 cases were a common set between the classification and segmentation steps. All images utilized in this study had corresponding liver segmentation generated by image analysts as the ground truth (Table 1).

The PLD dataset includes patients with a mean age of 57 years (range: 24 to 87), with males comprising 55% and females comprising 45% of the cohort. Regarding racial distribution, 93% of the patients identify as White, while the remaining patients belong to other races. The scanning manufacturers used in the dataset are as follows: 77% of the scans were conducted using Siemens scanners (Munich, Germany), 17% were performed using GE scanners (Boston, United States), and 6% were conducted using Toshiba scanners (Tokyo, Japan). In terms of scan parameters, 53% of the scans have a slice thickness of 5 mm, 27% have a slice thickness of 2.5 mm, and 13% have a slice thickness of 3 mm. The mean pixel size for the dataset is 0.76 mm (standard deviation: 0.0818).

The MCC cohort consists of patients with a mean age of 67 years (range: 26 to 93), with males comprising 57% and females comprising 43%. 97% of the patients identify as White. Regarding the scanning equipment used, 91% of the CT scans in this dataset were acquired using Siemens CT scanners, while the remaining scans were conducted using GE equipment. This information highlights the dominant presence of Siemens scanners in the dataset. Analyzing the scan parameters, it was observed that 50% of the scans have a slice thickness of 5 mm, indicating the level of detail captured in the images. Additionally, 46% of the scans have a slice thickness of 3 mm. These values provide insights into the variation in slice thickness within the dataset. The mean pixel size for the MCC dataset is recorded as 0.78 mm (standard deviation: 0.0813).

### Methods

Fig. 1a illustrates the overarching structure of our experimental design. Initially, we employ a deep neural network classifier as the initial stage of our approach. Its purpose is to classify patient CT scans into either the PLD or MCC class. The output of this classifier plays a crucial role in determining the subsequent segmentation model to be employed. To be more precise, for each pathology under consideration (PLD or MCC), we have constructed a model utilizing the U-Net architecture [12].

The workflow shown in Fig. 1b outlines the process of the application of a generic segmentation model that has been trained on both PLD and MCC data.

**Classification**

For the classification part of our system, the following models were investigated: ResNet[13], DenseNet[14], and EfficientNet[15]. 300 PLD scans and 300 MCC scans were used to create the training set. 450 scans were randomly selected for the training and 150 scans for the validation. 50 scans reserved from each of the PLD and MCC cohorts were used to create the test set.

The CT images were preprocessed before being used to train the classification models, as illustrated in Fig. 2. First, a CT window level of 180 HU (Hounsfield Unit) with a width of 440 HU was applied to enhance the soft tissue contrast. Next, the orientation of the CT images was standardized by rotating to orientation LPS (stands for "Left, Posterior, and Superior") for consistency. Then, images were resized to a uniform size of $128 \times 128 \times 128$ pixels to reduce GPU memory usage and computational complexity. The resizing was performed using box interpolation to preserve the overall structure and texture of the image while reducing its size. Finally, data augmentation techniques including random rotation and flipping were applied to improve the model's generalization performance and reduce the risk of overfitting the training data.

The classification models of ResNet, DenseNet, and EfficientNet families were trained using the same pre-processed dataset and identical settings to ensure a fair comparison and selection. A total of 300 epochs was used in all the model training. A batch size of 8 was used for each epoch. The cross-entropy loss and Adam optimizer[16] were configured for all the classification training models. In addition, a learning rate finder, which gradually increased the learning rate in a pre-training run to determine an optimal learning rate, was also implemented for all the model training. The performance of these models was compared on the validation set using key metric AUC (Area Under ROC curve). The DenseNet-121 model, which had the highest AUC, was selected as the best-performing classification model (Table 2). The confusion matrix for each model was analyzed to identify the strengths and weaknesses of each model in correctly classifying different categories. Metrics of accuracy, F1 score, precision, and sensitivity were used for sanity check and were included in Table 1 in the supplement material.

Furthermore, the selected DenseNet-121 classification model, with an optimal learning rate 1e-5 determined by the learning rate finder, was tuned based on the validation set to adapt to the PLD and MCC liver segmentation task. The tuning had two key elements. First, dropout was added to prevent the model from overfitting. A search based on the validation set was conducted and a dropout rate of 0.2 was found to provide the best classification results while keeping the model from overfitting. Second, class weighting was applied to the DenseNet-121 model. A weighting factor of 4 was assigned to the PLD class after conducting a search using the validation set. This enabled the DenseNet-121 model to enhance its learning of PLD cases, resulting in improved accuracy during classification, as observed through the analysis. The benefits of class weighting are elaborated in the discussion section.

**Segmentation**

The 3D nnU-Net model[17] was used for this study. The nnU-Net Python package version 1.7.0 was utilized to train the model. The preprocessing of the images and masks was automatically handled by the Python package (including resampling, intensity normalization, augmentation, etc.). 300 PLD scans and 300 MCC scans were used to train the PLD and MCC segmentation models, respectively. Additionally, a combination of both 300 PLD scans and 300 MCC scans was used to train the generic segmentation model. A test set consisting of 50 PLD and 50 MCC scans was used to evaluate the performance of both segmentation approaches. The segmentation test set was the same as that used for classification.

The training process involved 5-fold cross-validation for all the segmentation models, with each fold trained for 1000 epochs. The final model was an ensemble of five models obtained from the 5-fold training process. The training procedure was executed on Nvidia A100 GPUs with 80 GB of memory, utilizing a high-performance computing environment. The entire training process was completed in approximately one day.

**Metrics**

Multiple metrics were calculated to evaluate the performance of the classification model and the segmentation model, with a primary emphasis on two metrics. To evaluate the classifier's performance in classifying PLD and MCC pathology, *Area Under ROC curve (AUC)* was used as the key metric in selecting the best classification model in this study due to its ability to tell how well the model can separate the two classes. To evaluate the segmentation results, *Dice similarity Coefficient (Dice)* was used as the main measurement metric. More metrics to evaluate the segmentation results and the definition of classification and segmentation metrics are included in the supplement material.

**Statistical analysis**

Nonparametric Wilcoxon signed-rank tests were used across all the statistical analyses in the results. 95% confidence interval (CI) based on statistically significant evidence at $\alpha = 0.05$ was established for all the statistical analyses. All statistical analyses were performed with Python (version 3.8.13) and the Python SciPy library (version 1.8.1). In overview, the following statistical analyses were performed to investigate and compare the results (Table 2 in the supplement material).

Firstly, a non-parametric Wilcoxon signed-rank test was conducted to compare the segmentation output of the pipeline with that of the single generic model on the 100 test scans. The null hypothesis assumed that the two Dice distributions were the same, while the alternative hypothesis considered them to be different. Similarly, another test was conducted to compare the pipeline segmentation output with the optimal output.

Additionally, for a more detailed comparison between the pipeline segmentation output and the single model output, non-parametric Wilcoxon signed-rank tests were performed on the test scans within each classification category, namely, "PLD->PLD," "PLD->MCC," "MCC->MCC," and "MCC->PLD."[1]

Furthermore, to investigate whether correct classification leads to improvement in the adaptive segmentation pipeline, a non-parametric Wilcoxon signed-rank test was executed by comparing 50 PLD test scans segmented with the correct model against those segmented with the incorrect model. The same test was conducted for the 50 MCC test scans.

# Results

**Classification results**

100 test scans were used to evaluate the performance of the tuned DenseNet-121 model. For the 50 scans of PLD, 43 were correctly classified as containing PLD pathology, and 7 were misclassified as MCC. For the 50 scans of MCC, 46 were correctly classified as containing MCC pathology and 4 were misclassified as PLD pathology. The tuned

---

[1] "PLD->PLD" represented the PLD scans classified as PLD class and segmented with the PLD-trained U-Net model; "PLD->MCC" represented the PLD scans classified as MCC class and segmented with the MCC-trained U-Net model; same for "MCC->MCC" and "MCC->PLD".

DenseNet-121 model achieved an AUC of 0.956, an accuracy of 89.0%, and an F1 score of 89.0% on the test set. It reached a precision of 91.5% and sensitivity of 86.0% for the classification of the PLD class and a precision of 86.8% and sensitivity of 92.0% for the MCC class.

**Interpretability of the classification model**

To interpret the output of the tuned DenseNet-121 model, occlusion sensitivity maps were used. Fig. 3 illustrates an example from a PLD patient scan and an example from a MCC patient scan. Regions where the model's prediction was most sensitive to the changes or occlusions in the image, were displayed in red. For all the correctly predicted PLD and MCC scans, their occlusion sensitivity maps displayed a high sensitivity on the normal-appearing liver parenchyma. This suggested that the classification model depended on the normal-appearing liver parenchyma in the images to make decisions about whether it was a PLD scan or an MCC scan.

**Segmentation results**

The results of the adaptive segmentation pipeline were evaluated after the training of the DenseNet-121 model and U-Net segmentation models. Fig. 4 illustrates examples of the adaptive segmentation pipeline output on the PLD and MCC patient scans. In Fig. 5, the Dice comparison between the pipeline segmentation output, the optimal segmentation output, and the generic single U-Net segmentation output is provided. The optimal segmentation output refers to the segmentation results where all 100 PLD and MCC test scans were directed to the correct segmentation models using ground truth labels (equivalent to a 100% correct classifier in the pipeline). No statistically significant difference was found using the Wilcoxon signed-rank test between the pipeline segmentation output (average Dice of 0.971) and the optimal output (average Dice of 0.971) (n = 100, p-value = 0.131). However, a statistically significant difference was found using the Wilcoxon signed-rank test between the pipeline segmentation output (average Dice of 0.971) and the single segmentation model output (average Dice of 0.964) (n = 100, p-value << 0.001), suggesting the pipeline segmentation output performs better overall than the generic single model segmentation output despite the occasional pathology classification errors (Table 3).

Fig. 6 depicts the boxplot of the adaptive segmentation pipeline output on the test set, grouped by the real label of the scan and the segmentation model used. They are the same four categories presented in the classification results. A comparison of the Dice output between the segmentation pipeline and single segmentation model in each situation is presented in Table 3:

For the PLD test scans that were correctly classified and segmented, a statistically significant difference was found using the Wilcoxon signed-rank test between the pipeline segmentation output (average Dice of 0.962) and the single segmentation model output (average Dice of 0.956) (n = 43, p-value << 0.001), with observed average Dice coefficient suggesting the pipeline segmentation output performs better than the generic single model segmentation output; for the incorrectly classified and segmented PLD test cases, a statistically significant difference was found using the Wilcoxon signed-rank test between the pipeline segmentation output (average Dice of 0.964) and the single segmentation model output (average Dice of 0.967) (n = 7, p-value = 0.0156), with observed average Dice coefficient suggesting the pipeline segmentation output performs slightly worse than the generic single model segmentation output; for the MCC test scans that were correctly classified and segmented, a statistically significant difference was found using the Wilcoxon signed-rank test between the pipeline segmentation output (average Dice of 0.980) and the single segmentation model output (average Dice of 0.970) (n = 46, p-value << 0.001), with observed average Dice coefficient suggesting the pipeline segmentation output performs better than the generic single model segmentation output; for the MCC test scans that were incorrectly classified and segmented, no statistically significant difference was found using the Wilcoxon signed-rank test between the pipeline segmentation output (average Dice of 0.974) and the single segmentation model output (average Dice of 0.984) (n = 4, p-value = 0.125).

In summary, although performance was worse when the classifier result was incorrect and the wrong pathology-specific model was used, the number of classifier errors was low enough that the pipeline model overall performed better on the entire set of test cases.

## Discussion

Our adaptive segmentation pipeline showed improved results in the accuracy and consistency of liver segmentation in PLD and MCC CT scans, compared to generic models trained on a mixture of the two pathologies. We hope that it provides a more effective and adaptable approach to provide a comprehensive and reliable solution for segmenting organs affected by diverse pathologies. Improvements in the reliability and generalizability of automated segmentation algorithms can reduce the time and resources required for manual classification and segmentation. This pipeline approach could be a valuable tool for research and clinical practice.

While building the adaptive segmentation pipeline, there were several considerations. The first was to have an accurate classification model. Fig. 7 shows the boxplot of the Dice of 50 PLD test scans and 50 MCC test scans segmented with the correct or incorrect models. A nonparametric Wilcoxon signed-ranks test on the Dice of "PLD->PLD" (average 0.963) and "PLD->MCC" (average 0.775) suggested a difference in the segmentation accuracy (n = 50, p-value << 0.001), with observed Dice coefficients indicating the "PLD->PLD" resulted in higher accuracy (Table 4). Similarly, the nonparametric Wilcoxon signed-ranks test on the Dice of "MCC->MCC" (average 0.980) and "MCC->PLD" (average 0.937) suggested a difference in the segmentation accuracy (n = 50, p-value << 0.001), with observed Dice coefficients indicating the "MCC->MCC" resulted in higher accuracy (Table 4). Thus, it is important to have an accurate classifier in the adaptive pipeline so that the CT scans can be segmented with the "correct" model.

The second consideration was to find the optimal balance between the classification model complexity and generalization performance. Increasing the complexity of the classification model such as using more layers, often resulted in improved training performance but decreased generalization performance due to overfitting of the model. To overcome this, the dropout technique was used to prevent overfitting and improve generalization performance. Of note, the optimal dropout rate can depend on the image size used, available GPU memory, and the particular segmentation task.

Third, the imbalance of classifying and segmenting PLD and MCC scans was considered. Although the number of CT scans for training is the same for each class, the classification model has different levels of difficulty in classifying one pathology from the other. In this study, the classification model tended to make more mistakes by classifying the PLD scans as MCC. Similarly, the segmentation model had different abilities for segmenting the PLD and MCC scans with the PLD scans being more challenging to segment, likely due to a more lobulated liver contour. For example, in Table 4, the "PLD->PLD" category had a lower Dice average (0.963), than the "MCC->MCC" category (0.980). The imbalancing nature of this task would lead to worse segmentation results for misclassified PLD scans than for misclassified MCC scans, as shown in Fig. 7. Therefore, it was beneficial to apply class weighting to weight the PLD class more during classification so that fewer PLD scans were misclassified as MCC. Figure 6 demonstrates the improved segmentation performance of class weighting in the Dice score distribution.

Lastly, we analyzed the PLD and MCC scans with low Dice scores in the adaptive segmentation pipeline results (the ones outside the error bar in the pipeline output in Fig. 5). Although the livers had large degree of anatomic distortion, these scans were correctly classified by the adaptive segmentation pipeline (Table 5). Their segmentation results are better than the ones from the single segmentation model. No systematic reason of misclassification was identified.

**Limitations**

While our proposed adaptive segmentation pipeline has shown improved segmentation results, it also has limitations. Firstly, our proposed pipeline requires a classification step in its initial phase, which poses a challenge to the training of an accurate classifier. Secondly, our two-step process may demand more effort and computing resources compared to the generic approach. Thirdly, the intricate difficult cases rely on a further update of the segmentation model in the pipeline to improve the segmentation accuracy in those cases. Fourthly, due to limited data availability, only a limited number of PLD and MCC pathology CT scans (350 scans each) are used for this feasibility study, and a binary classifier is trained in this pipeline. To enable broader classification possibilities, a multi-class classifier is necessary for the first classification step.

## Conclusion

In this research, we have introduced an adaptive liver segmentation pipeline designed to effectively segment images of different pathologies. The pipeline leverages a deep learning classifier to categorize the images and then directs them to the appropriate models for segmentation. The results of our study show a significant statistical improvement in segmentation compared to using a generic single segmentation model trained on a pooled dataset of both pathologies. This approach could integrate specialized segmentation models into a unified workflow, facilitating the segmentation of diverse images and enabling a potential solution for clinical applications.

## Conflict of Interest

The authors declare that they have no conflict of interest.

**Table 1.** Summary of the Polycystic Liver Disease (PLD) and Metastases from Colorectal Cancer (MCC) data used for training, validation, and testing.

|  | **Polycystic Liver Disease** | **Metastases from Colorectal Cancer** |
|---|---|---|
| **Number of CT scans** | 350 | 350 |
| **Male-to-female ratio** | 1.22 | 1.33 |
| **Age, mean(min:max)** | 57 (24:87) | 67(26:93) |
| **Race** | 93% White<br>7% Other | 97% White<br>3% Other |
| **Manufacturer model** | 77% Siemens<br>17% GE<br>6% Toshiba | 91% Siemens<br>9% GE |
| **Slice thickness** | 53% 5 mm<br>13% 3 mm<br>27% 2.5 mm<br>7% other | 50% 5 mm<br>46% 3 mm<br>3% other |
| **Pixel size, mean(std. deviation)** | 0.76 mm (0.0818) | 0.78 mm (0.0813) |

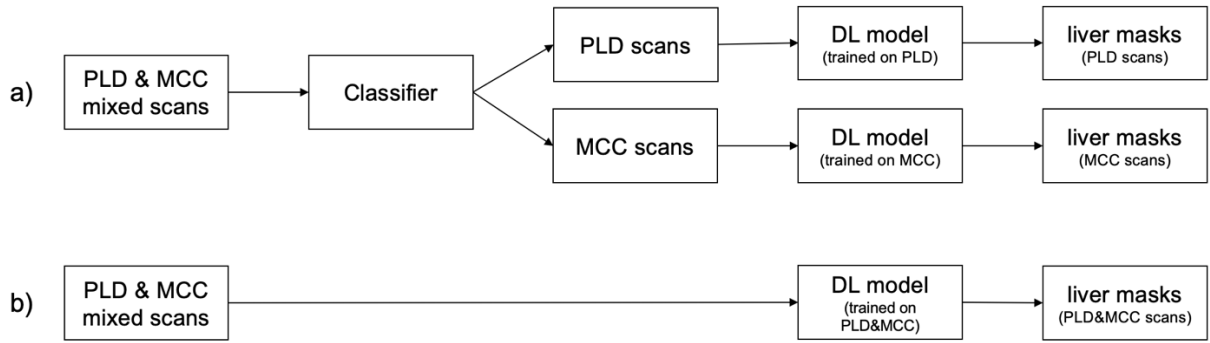

**Fig. 1** a) The workflow of the adaptive segmentation pipeline and b) the workflow of the generic segmentation model on the PLD and MCC CT scans. (PLD: Polycystic Liver Disease, MCC: Metastases from Colorectal Cancer, DL: Deep Learning)

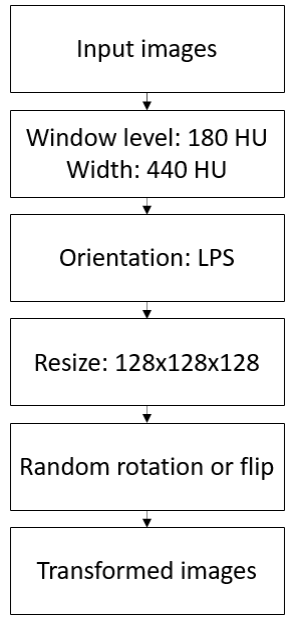

**Fig. 2** Preprocessing steps of the PLD and MCC CT images for the classification model training. Random rotations are available in 90, 180, and 270 degrees.

**Table 2.** Performance of various classification models evaluated on the PLD and MCC validation set (150 scans were used for the validation out of 600 training scans) using Area Under ROC curve (AUC).

| Model | DenseNet (DN) | | | | EfficientNet (EN) | | | | ResNet (RN) | | |
|---|---|---|---|---|---|---|---|---|---|---|---|
| | DN-121 | DN-169 | DN-201 | DN-264 | EN-b0 | EN-b1 | EN-b2 | EN-b3 | RN-10 | RN-18 | RN-34 |
| AUC | *0.970* | 0.969 | 0.964 | 0.966 | 0.925 | 0.929 | 0.968 | 0.962 | 0.936 | 0.936 | 0.961 |

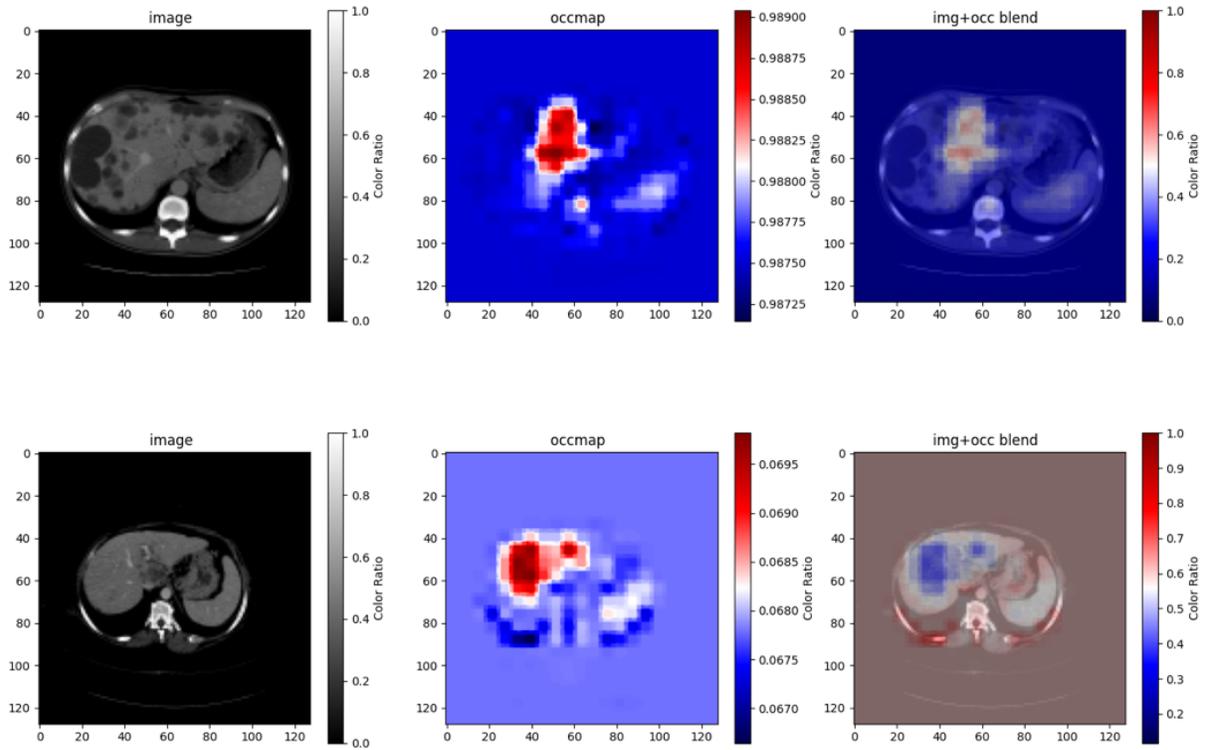

**Fig. 3** Illustration of occlusion sensitivity maps for the correctly classified PLD (top row) and MCC (bottom row) scans in the test set. The first column displays the original image; the second column displays the occlusion sensitivity map; and the third column displays the blend of the image and the occlusion sensitivity map.

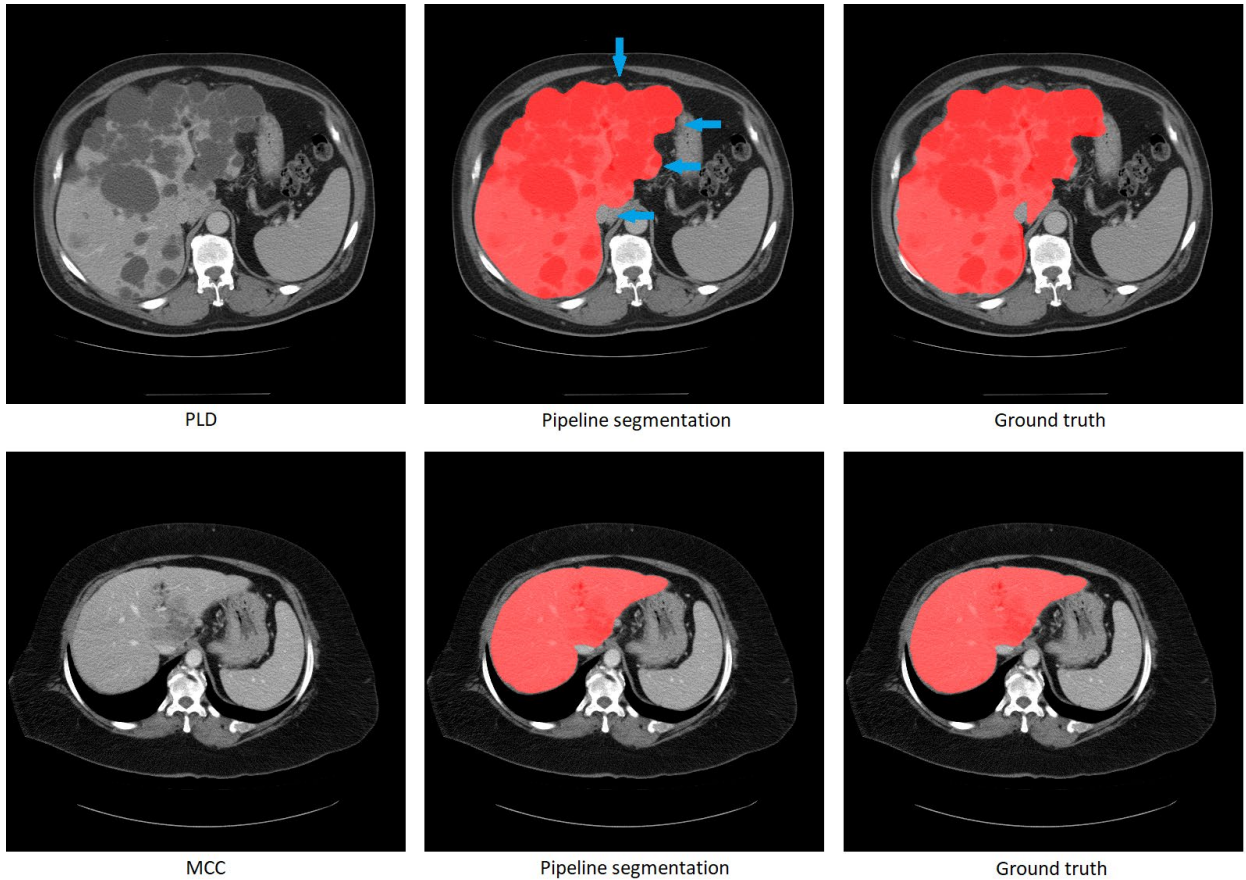

**Fig. 4** Illustration of the pipeline segmentation results for a PLD patient (top row) and an MCC patient (bottom row). The first column displays the original image; the second column showcases the pipeline's segmentation outcome; and the third column presents manual liver segmentations by analysts. Both the PLD and MCC image were correctly classified and segmented using the pathology-specific segmentation model. The blue arrows point to areas where the pipeline segmentation results are different from the ground truth.

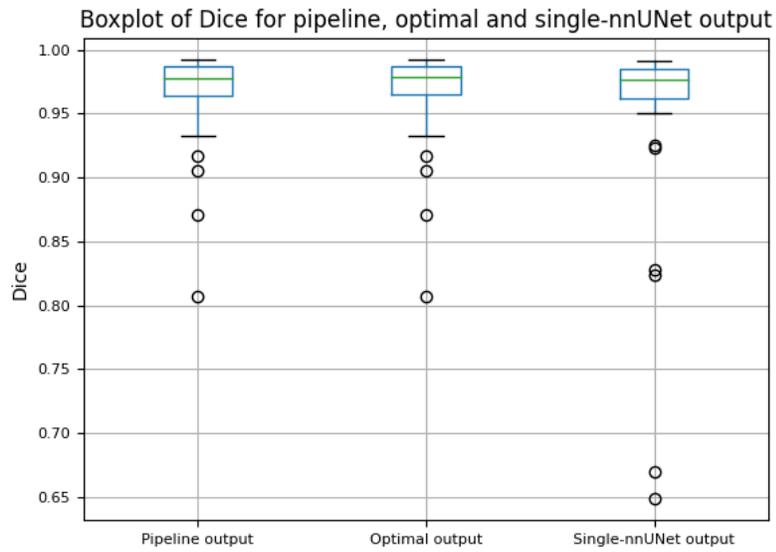

**Fig. 5** Boxplot comparing Dice Similarity Coefficient (Dice) of the adaptive segmentation pipeline output, optimal segmentation output, and single nnU-Net segmentation output. The optimal segmentation output refers to the segmentation results where all PLD and MCC test scans were directed to the correct segmentation models using ground truth labels.

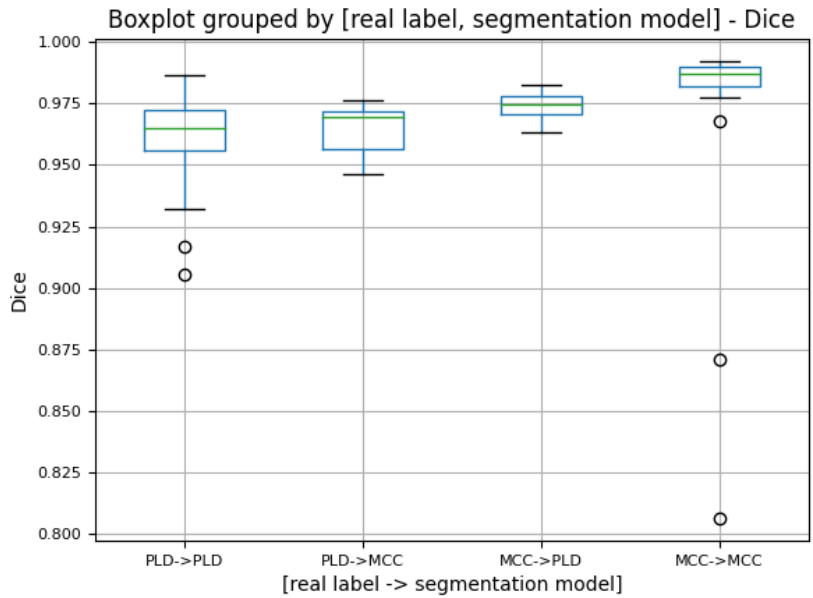

**Fig. 6** Boxplot of Dice Similarity Coefficient (Dice) of the adaptive segmentation pipeline output grouped by the real label of the scan and the segmentation model used for that scan. For the 50 PLD and 50 MCC scans, 43 PLD scans were classified as PLD, 7 PLD scans were classified as MCC, 4 MCC scans were classified as PLD and 46 MCC scans were classified as MCC. These scans categorized here are the same as the ones presented in the classification results.

**Table 3.** Nonparametric Wilcoxon signed-rank test results on the Dice output of pipeline vs optimal, pipeline vs single model and different categories of pipeline vs. single model.

| Samples | Number of scans | Pipeline Dice avg. | Optimal Dice avg. | p-value |
|---|---|---|---|---|
| All 100 test scans | 100 | 0.971 | 0.971 | 0.131 |

| Samples | Number of scans | Pipeline Dice avg. | Single model Dice avg. | p-value |
|---|---|---|---|---|
| All 100 test scans | 100 | *0.971* | 0.964 | $1.57 \times 10^{-5}$ |

| Samples (Real label ->segmentation model) | Number of scans | Pipeline Dice avg. | Single model Dice avg. | p-value |
|---|---|---|---|---|
| PLD->PLD | 43 | *0.962* | 0.956 | $4.84 \times 10^{-4}$ |
| PLD->MCC | 7 | 0.964 | *0.967* | 0.0156 |
| MCC->PLD | 4 | 0.974 | 0.984 | 0.125 |
| MCC->MCC | 46 | *0.980* | 0.970 | $2.15 \times 10^{-8}$ |

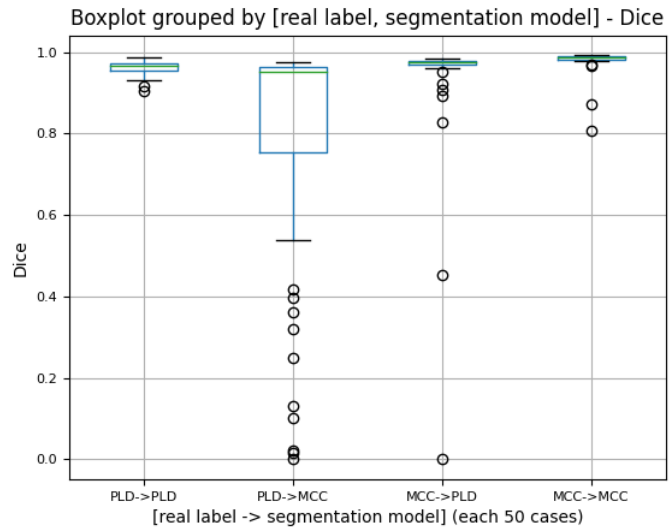

**Fig. 7** Boxplot of Dice Similarity Coefficient (Dice) of the test scans (50 PLD and 50 MCC) segmented with the "correct" or "incorrect" U-Net model. For example, "PLD->PLD" represents the Dice distribution of 50 PLD scans all segmented with the PLD-trained U-Net model; "PLD->MCC" represents the Dice distribution of 50 PLD scans all segmented with the MCC-trained U-Net model.

**Table 4.** Nonparametric Wilcoxon signed-rank tests on the 50 PLD test scans correctly segmented vs incorrectly segmented, and on the 50 MCC scans correctly segmented vs incorrectly segmented.

| Samples | Number of scans | PLD->PLD Dice avg. | PLD->MCC Dice avg. | p-value |
|---|---|---|---|---|
| 50 PLD test scans | 50 | *0.963* | 0.775 | $1.13 \times 10^{-8}$ |

| Samples | Number of scans | MCC->MCC Dice avg. | MCC->PLD Dice avg. | p-value |
|---|---|---|---|---|
| 50 MCC test scans | 50 | *0.980* | 0.937 | $1.23 \times 10^{-9}$ |

**Table 5.** Comparison of the difficult-to-segment PLD and MCC scans in the test set when segmented with the adaptive segmentation pipeline vs. the single segmentation model.

| Patient scan | Real label | Predicted label | Pipeline Dice | Single model Dice |
|---|---|---|---|---|
| 1 | MCC | MCC | *0.807* | 0.649 |
| 2 | MCC | MCC | *0.871* | 0.670 |
| 3 | PLD | PLD | *0.905* | 0.824 |
| 4 | PLD | PLD | *0.941* | 0.827 |

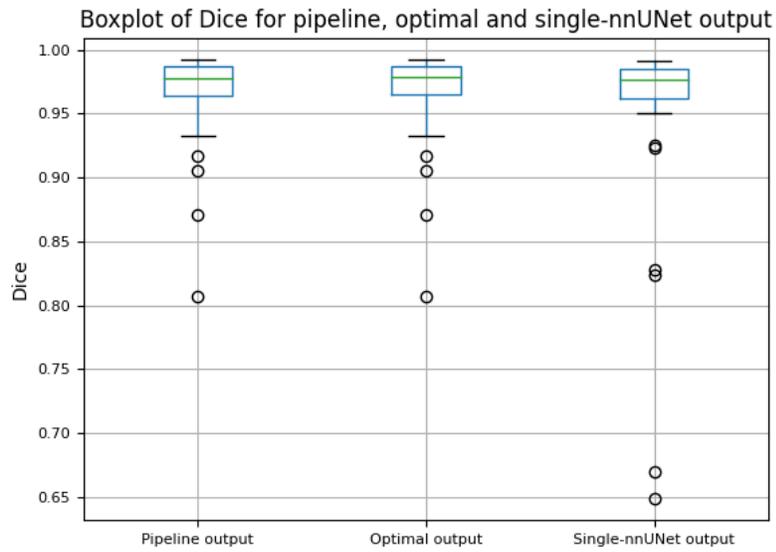

**Fig. 5** Boxplot comparing Dice Similarity Coefficient (Dice) of the adaptive segmentation pipeline output, optimal segmentation output, and single nnU-Net segmentation output. The optimal segmentation output refers to the segmentation results where all PLD and MCC test scans were directed to the correct segmentation models using ground truth labels.

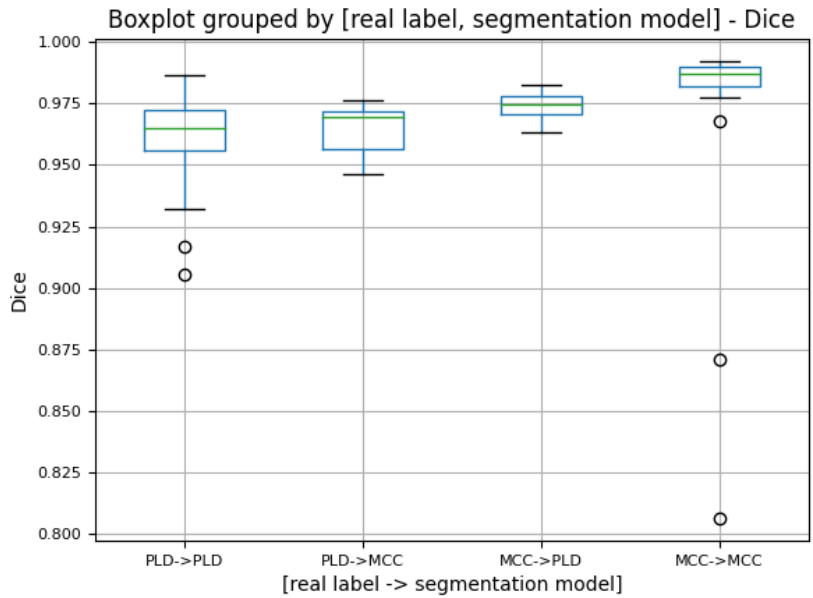

**Fig. 6** Boxplot of Dice Similarity Coefficient (Dice) of the adaptive segmentation pipeline output grouped by the real label of the scan and the segmentation model used for that scan. For the 50 PLD and 50 MCC scans, 43 PLD scans were classified as PLD, 7 PLD scans were classified as MCC, 4 MCC scans were classified as PLD and 46 MCC scans were classified as MCC. These scans categorized here are the same as the ones presented in the classification results.

**Table 3.** Nonparametric Wilcoxon signed-rank test results on the Dice output of pipeline vs optimal, pipeline vs single model and different categories of pipeline vs. single model.

| Samples | Number of scans | Pipeline Dice avg. | Optimal Dice avg. | p-value |
|---|---|---|---|---|
| All 100 test scans | 100 | 0.971 | 0.971 | 0.131 |

| Samples | Number of scans | Pipeline Dice avg. | Single model Dice avg. | p-value |
|---|---|---|---|---|
| All 100 test scans | 100 | *0.971* | 0.964 | $1.57 \times 10^{-5}$ |

| Samples (Real label ->segmentation model) | Number of scans | Pipeline Dice avg. | Single model Dice avg. | p-value |
|---|---|---|---|---|
| PLD->PLD | 43 | *0.962* | 0.956 | $4.84 \times 10^{-4}$ |
| PLD->MCC | 7 | 0.964 | *0.967* | 0.0156 |
| MCC->PLD | 4 | 0.974 | 0.984 | 0.125 |
| MCC->MCC | 46 | *0.980* | 0.970 | $2.15 \times 10^{-8}$ |

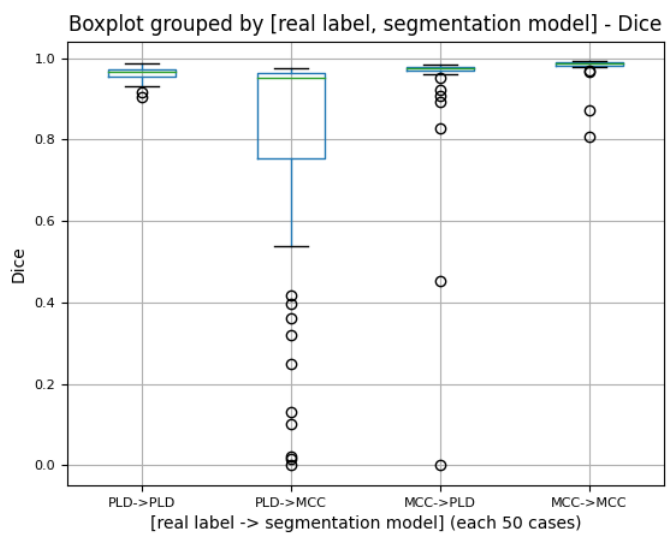

**Fig. 7** Boxplot of Dice Similarity Coefficient (Dice) of the test scans (50 PLD and 50 MCC) segmented with the "correct" or "incorrect" U-Net model. For example, "PLD->PLD" represents the Dice distribution of 50 PLD scans all segmented with the PLD-trained U-Net model; "PLD->MCC" represents the Dice distribution of 50 PLD scans all segmented with the MCC-trained U-Net model.

Table 4. Nonparametric Wilcoxon signed-rank tests on the 50 PLD test scans correctly segmented vs incorrectly segmented, and on the 50 MCC scans correctly segmented vs incorrectly segmented.

| Samples | Number of scans | PLD->PLD Dice avg. | PLD->MCC Dice avg. | p-value |
|---|---|---|---|---|
| 50 PLD test scans | 50 | *0.963* | 0.775 | $1.13 \times 10^{-8}$ |

| Samples | Number of scans | MCC->MCC Dice avg. | MCC->PLD Dice avg. | p-value |
|---|---|---|---|---|
| 50 MCC test scans | 50 | *0.980* | 0.937 | $1.23 \times 10^{-9}$ |

**Table 5.** Comparison of the difficult-to-segment PLD and MCC scans in the test set when segmented with the adaptive segmentation pipeline vs. the single segmentation model.

| Patient scan | Real label | Predicted label | Pipeline Dice | Single model Dice |
|---|---|---|---|---|
| 1 | MCC | MCC | *0.807* | 0.649 |
| 2 | MCC | MCC | *0.871* | 0.670 |
| 3 | PLD | PLD | *0.905* | 0.824 |
| 4 | PLD | PLD | *0.941* | 0.827 |